# Observation of chiral currents at the magnetic domain boundary of a topological insulator

Y. H. Wang[1,2*], J. R. Kirtley[1], F. Katmis[3,4], P. Jarillo-Herrero[4], J. S. Moodera[3,4], K. A. Moler[1,2*]

1. Department of Physics and Applied Physics, Stanford University, Stanford, CA 94305, USA

2. Stanford Institute for Materials and Energy Sciences, SLAC National Accelerator Laboratory, Menlo Park, CA 94025, USA

3. Francis Bitter Magnet Lab and Plasma Science and Fusion Center, Massachusetts Institute of Technology, Cambridge, MA 02139, USA

4. Department of Physics, Massachusetts Institute of Technology, Cambridge, MA 02139, USA

* To whom correspondence and requests for materials should be addressed. Email: wangyhv@stanford.edu (Y.H.W.); kmoler@stanford.edu (K.A.M.)

**A magnetic domain boundary on the surface of a three-dimensional topological insulator is predicted to host a chiral edge state, but direct demonstration is challenging. Here, we used a scanning superconducting quantum interference device to show that current in a magnetized EuS/$Bi_2Se_3$ heterostructure flows at the edge when the Fermi level is gate-tuned to the surface band gap. We further induced micron-scale magnetic structures on the heterostructure, and detected a chiral edge current at the magnetic domain boundary. The chirality of the current was determined by magnetization of the surrounding domain and its magnitude by the local chemical potential rather than the applied current. Such magnetic structures, provide a platform for detecting topological magnetoelectric effects and may enable progress in quantum information processing and spintronics.**

The metallic surface of a three-dimensional topological insulator (3D-TI) is protected by time-reversal symmetry (TRS). Breaking TRS opens a band gap on the surface Dirac cone and transforms it into a Chern insulator (*1-4*). TRS-broken surface states are predicted to exhibit topological magneto-electric effects (*1*), and, when coupled with a superconductor, Majorana fermions (*5-7*). Just as the surface Dirac cone is a signature of the non-trivial topological bulk band structure of a time-reversal invariant 3D-TI, bulk-boundary correspondence dictates that the TRS-broken surface states with a nonzero Chern number manifest as a gapless chiral edge state (CES) at its boundary (*1*).

In the special case where the boundary is the edge of the sample surface, a CES along the edge leads to a quantized Hall conductance equal to $e^2/h$, where $e$ is the electron charge and $h$ is the Planck constant, even at zero magnetic field. This quantized anomalous Hall conductance was observed only in a 3D-TI doped with a high concentration of magnetic impurities to break TRS, with the measurements performed at very low temperatures to achieve ballistic transport between contacts (*8, 9*). More generally, a CES theoretically should exist at a magnetic domain boundary (*1, 10*), which does not need to be the physical boundary of the system. In this case, the presence of a CES changes only the local conductivity and therefore does not contribute to the conductance of the system (*11*). The CES at a magnetic domain boundary can be used to investigate one-dimensional quantum transport without edge effects (*12-14*), to induce a parity anomaly (*4*), or to realize magnetically defined quantum bits (*15*).

Inducing magnetism on the surface of a TI through proximity to a ferromagnet provides an alternative strategy for breaking the TRS on the surface states without disrupting the bulk. Previous bulk magnetic and transport measurements (*16*) revealed some evidence of enhanced magnetization at the interface of a heterostructure composed of EuS, a known ferromagnetic insulator (FMI) (*17-19*), and $Bi_2Se_3$, a prototypical TI (*20-23*) (Fig. 1A). For our experiments, a Hall bar was etched through the bilayer using a shadow mask (Fig. 1B) for transport measurements. The Curie temperature of the EuS(10nm)/Bi$_2$Se$_3$(5nm) heterostructure was previously found by magneto-resistance and bulk magnetic measurement to be approximately 15 K (*16, 24*), comparable to that of bulk EuS (*17*).

We employed scanning superconducting quantum interference device (SQUID) microscopy (Fig. 1A), a sensitive probe for detecting magnetic flux from magnetic domains or current on the mesoscopic scale (*25, 26*), to search for the chiral edge states in a EuS/Bi$_2$Se$_3$ heterostructure (Fig. 1B). The pickup loop (Fig. 1A) was integrated into a two-junction SQUID that converts the flux through the loop ($\Phi$) into a voltage signal (*27-29*). Flowing a current ($I_F$) through the field coil (Fig. 1A) provided a local magnetic field for either susceptometry measurement ($d\Phi/dI_F$) or to manipulate magnetic domain structures in our bilayers (Fig. 1A). Current magnetometry ($\Phi'_I/I_{AC}$) was performed simultaneously with direct current (DC) magnetometry by measuring the component of the flux through the pickup loop that is locked to the frequency and in phase ($\Phi'_I$) with the alternating current (AC) source-drain bias current ($I_{AC}$).

A typical magnetometry micrograph (Fig. 1C) of the sample under zero field cooling displayed micron-scale patches of magnetic domains at the base temperature of 4 K. Such magnetic features disappeared at 19 K (Fig. 1C, inset), consistent with the Curie temperature of these samples (*16, 24*). The etched area of the Hall bar showed zero magnetization (Fig. 1C) and susceptibility (Fig. 1D). The much-reduced susceptibility of the film at 21 K (Fig. 1D inset) was also consistent with its ferromagnetic nature. To break the TRS at the EuS/Bi$_2$Se$_3$ interface with magnetization, we applied a uniform out-of-plane magnetic field of 30 G while cooling the sample from 18 K to 5 K *(30)*. After field cooling, the magnetic structure close to the edge of the Hall bar, as determined from scanning susceptometry (Fig. 2A) showed a change of magnetization from the film side to the substrate side; this indicated that the out-of-plane magnetic field induced, on average, a mixture of out-of-plane and in-plane remnant magnetization (*31*). The lateral variation of magnetization along the edge may be caused by the inhomogeneity of the film and some domain structure (*31*). The magnetic field from the magnetization is less than 1 G (*31*) and is too small to induce any observable Landau levels (*32*) at 4 K.

To determine how current flows around the edge of a Hall bar when it is magnetized, we performed simultaneously magnetometry and current magnetometry at various back-gate voltages $V_G$ (Fig. 2). Although the magnetometry images did not change as a function of $V_G$ (Fig. 2, A to C), current magnetometry images strongly depended on $V_G$ (Fig. 2, D to F). We found a current magnetometry pattern with both positive and negative polarity developing along the edge as the back-gate voltage was tuned to be more negative (Fig. 2D). The current magnetometry pattern reached a maximum at $V_G = -220$ V (Fig. 2E). The pattern became weaker with decreasing $V_G$ and completely disappeared at $V_G = -350$ V (Fig. 2F). The cross sections normal to the edge from these images (Fig. 2G) clearly exhibited a field profile consistent with an edge current developing when the Fermi level is gate-tuned to the surface band gap induced by magnetism (*25*). The extracted current density from the flux image at $V_G = -220$ V indicates that the edge current appears to be confined to the edge with a full-width half maximum width of 4.1 µm (Fig. 2H), which is likely resolution-limited for the 3 µm-diameter pickup loop at a scan height of ~1 µm in this particular measurement (*31*).

Having demonstrated the existence of edge states at the physical boundary of the heterostructure, we investigated whether there were any CESs at a magnetic domain boundary. We applied DC current to the field coil at 12 K while scanning the field coil in a square (Fig. 3A) to write a magnetic structure with two opposite out-of-plane magnetizations next to each other and away from the edge (*31*).

To investigate the current around this magnetic structure, we performed current magnetometry with source-drain voltage biased both ways (Fig. 3). The reversal of the bias voltage does not cause any change to the induced magnetic structure (Fig. 3, A and B). For the current flowing in a conventional metal, the direction of the current and therefore its magnetic field reverses sign when the source-drain bias is reversed (fig. S4). Such non-chiral current is present in our TI-FMI heterostructure: on the left side of the images in Fig. 3, C and D, in which average magnetization is zero (Fig. 3A) as well as in the image with $V_G = 0$ V (Fig. 3E), there is a linear background that switches sign when the bias current is switched. This is consistent with the out-of-plane magnetic field from a uniform plane current flowing horizontally (fig. S5A) and likely comes from the bottom surface or the bulk. The background current is caused by the gradient of chemical potential across the thickness of the $Bi_2Se_3$ layer even when the chemical potential on the top is tuned to inside the energy gap of the surface states by the back-gate voltage (*33*), as is also known to happen in two-dimensional quantum spin Hall systems (*25*).

In addition, sharp current flux features appeared that, in contrast to the background current, did not switch sign as the voltage bias was reversed (Fig. 3, C and D). These features, reminiscent of the flux generated by an edge current (Fig. 2E), appeared along the magnetic domain wall at $V_G = $ -220 V (Fig. 3, C and D) but were absent at $V_G = 0$ V (Fig. 3E), which is consistent with the gate dependence of the edge current (Fig. 2) and suggests their common origin from the states in the surface gap. The sign of these features is opposite (Fig. 3, C and D) along the left edges of the top and bottom domains with reversed magnetizations (Fig. 3A), indicating the chiral nature of the edge current. These chiral features in current magnetometry disappeared when the temperature was increased to 12 K, whereas the magnetization became weaker but was still present (fig. S6). This observation suggests that a strong magnetization is essential to the chiral current features and therefore only the top surface of $Bi_2Se_3$ is contributing to such features because exchange coupling from the FMI layer is short range (*34, 35*). In stark contrast, current magnetometry on a trivial semimetal - FMI bilayer showed no such chiral features around a magnetic structure at base temperature regardless of gating (fig. S4).

The average of the current magnetometry images with opposite source-drain bias cancelled out the non-chiral background current and clearly reveals the chiral features along the magnetic domain boundary in the shape of a square (fig. S3B). We find that the peaks in current magnetometry were approximately described using the out-of-plane magnetic field generated by thin current wires with finite width (*31*). The current density (Fig. 3F) extracted from the current magnetometry image (Fig. 3D) (*25*) confirmed this picture of chiral edge current surrounding the magnetic domain boundary with chirality determined by magnetization of the domain. Because the magnetization we induced with the field coil was below the saturation magnetization of similar films (*16*), there could be domains beyond the resolution of the pickup loop inside the magnetic structures (Fig. 3A). However, the chiral current from these smaller domains only reduces the average intensity of chiral features in current magnetometry in a similar fashion that these domains reduce the average magnitude of magnetization detected by the pickup loop (*31*).

To identify what determines chiral current intensity, we performed current magnetometry on the magnetic structures induced at various positions along the Hall bar, where the electrochemical potentials ($\mu$) differ (Fig. 4). Magnetic structures were induced at different locations along the Hall bar (Fig. 4A and fig. S12) by cooling the sample from 15 K while applying DC current to the field coil (*31*). The current magnetometry images (Fig. 4, B to G) correspond to such structures induced on the left of the Hall bar close to the source contact (Fig. 4, B and C), in the middle of the Hall bar (Fig. 4, D and E), and on the right of the Hall bar close to the drain contact (Fig. 4, F and G). Although the sign of the current features were the same for all configurations (at $V_G$ = -200 V) (Fig. 4, B to G), as expected from a chiral effect, the intensity was strongly position dependent. For the magnetic structure close to the source, the intensity of the current features was stronger when the voltage bias was on the source (Fig. 4B) than when the bias was on the drain (Fig. 4C). This behavior was reversed for the magnetic structure close to the drain (Fig. 4, F and G), whereas the middle structure displayed approximately equal intensity when the voltage bias was switched (Fig. 4, D and E).

Because the location of the application of the voltage bias does not change the total current flowing through the Hall bar, the position-dependent current flux intensity suggests that $\mu$ rather than the bias current $I_{AC}$ determined the chiral current intensity. $\mu$ at each location was linearly proportional to the position along the Hall bar because the Hall bar's length is much longer than the typical electron mean free path (Fig. 1B, fig. S12D). This dependence on $\mu$ is reminiscent of the chiral edge current in a quantum Hall state, where the domain boundary is the edge of the sample (*34*). Current flowing along each edge is proportional to the $\mu$ of the contact from which the charge is emitted. Meanwhile, the total bias current depends on the difference in $\mu$ between the source and drain contacts. If the voltages on the contacts are equal, then there will be no net current, but there will be a circulating edge current with a magnitude dependent on the source-drain voltage relative to the back-gate voltage (*11*).

In our TI-FMI heterostructure, the drop of $\mu$ across the magnetic structure was small compared to the drop of $\mu$ across source and drain (due to the ratio of their lengths) and therefore the local $\mu$ determined the magnitude of the chiral edge current. Because current magnetometry is a lock-in measurement (*31*), the measured magnetic response from the chiral current is proportional to the modulation of the local $\mu$ when the Fermi level is in the surface gap. When the domain boundary is close to the alternating AC voltage bias (Fig. 4, B and G), $\mu$ is more strongly modulated than when it is close to the ground (Fig. 4, C and F), yielding a stronger chiral current intensity in current magnetometry (Fig. 4H). The magnitude of the extracted current $I$ (fig. S5B) along the magnetic domain boundary (*31*) is in agreement with calculations of the current carried by one spin-polarized edge mode in the ballistic transport regime $I = e\mu/h$ (*11, 35, 36*).

Our results not only demonstrate the existence of CES at the magnetic domain boundary of a TI but also establish a versatile platform in scanning SQUID microscopy for imaging and manipulating broken TRS TI surface states on the mesoscopic scale. The broken TRS state and its chiral edge will be a playground for exploring interaction between TIs and FMIs (*37-40*).

Content of the Supplementary Online Material
- **I.** **Materials and Methods**
- **II.** **Supplementary Text**
- **III.** **Figs. S1 to S12**

## Acknowledgments:

This work is supported by FAME, one of six centers of STARnet, a Semiconductor Research Corporation program sponsored by MARCO and DARPA. The SQUID microscope and sensors used were developed with support from the NSF-sponsored Center for Probing the Nanoscale at Stanford, NSF-NSEC 0830228, from NSF IMR-MIP 0957616 and from the Department of Energy, Office of Science, Basic Energy Sciences, Materials Sciences and Engineering Division, under Contract DE-AC02-76SF00515. Y.H.W. is partially supported by the Urbanek Fellowship of the Department of Applied Physics at Stanford University. F.K., P. J-H. and J.S.M. would like to thank support by the MIT MRSEC through the MRSEC Program of the National Science Foundation under award number DMR-0819762. Partial support was provided by NSF (DMR-1207469), ONR (N00014-13-1-0301) (F.K. and J.S.M.) and by the DOE, Basic Energy Sciences Office, Division of Materials Sciences and Engineering, under Award No. DE-SC0006418 (F.K. and P. J-H.). We are grateful for the assistance from Gerald Gibson, Mark Ketchen, and Martin Huber on developing the SQUID sensors and from I. Sochnikov, E. Spanton, J. Palmstrom, and K. C. Nowack on the measurement, as well as for stimulating discussions with S.-C. Zhang, X.-L. Qi, J. Wang, D. Goldhaber-Gordon, and B. I. Halperin.

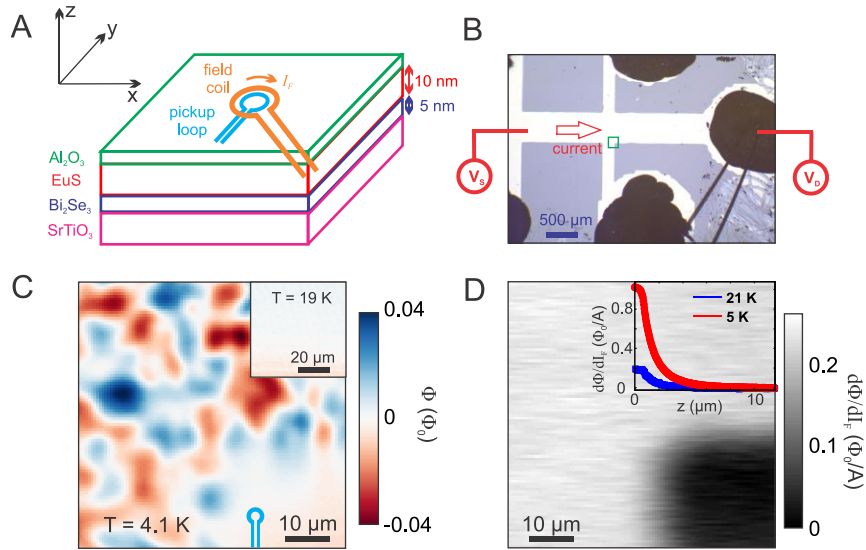

**Fig. 1 Scanning SQUID microscopy of a TI - FMI heterostructure shows micron-scale magnetic domains.** (**A**) Schematic of the EuS/Bi$_2$Se$_3$ bilayer nanostructure on the SrTiO$_3$ substrate under the pickup loop and field coil of a SQUID. (**B**) Optical image of the patterned bilayer film. The Hall bar has all the layers shown in (**A**) and the etched area only has the SrTiO$_3$ substrate. (**C**) Scanning SQUID flux (Φ) image of the green square area in (**B**) at 4.1 K after zero-field cooling and at 19 K (inset).). The pickup loop (cyan) is shown to scale. Φ$_0$ is the flux quantum. (**D**) Susceptometry (dΦ/dI$_F$) micrograph of the area in (**C**). Inset: susceptibility as a function of distance from the surface of the film at 5 K (red) and 21 K (blue). The lower right corner of (**C**) and (**D**), which corresponds to an etched area, shows small flux and susceptibility.

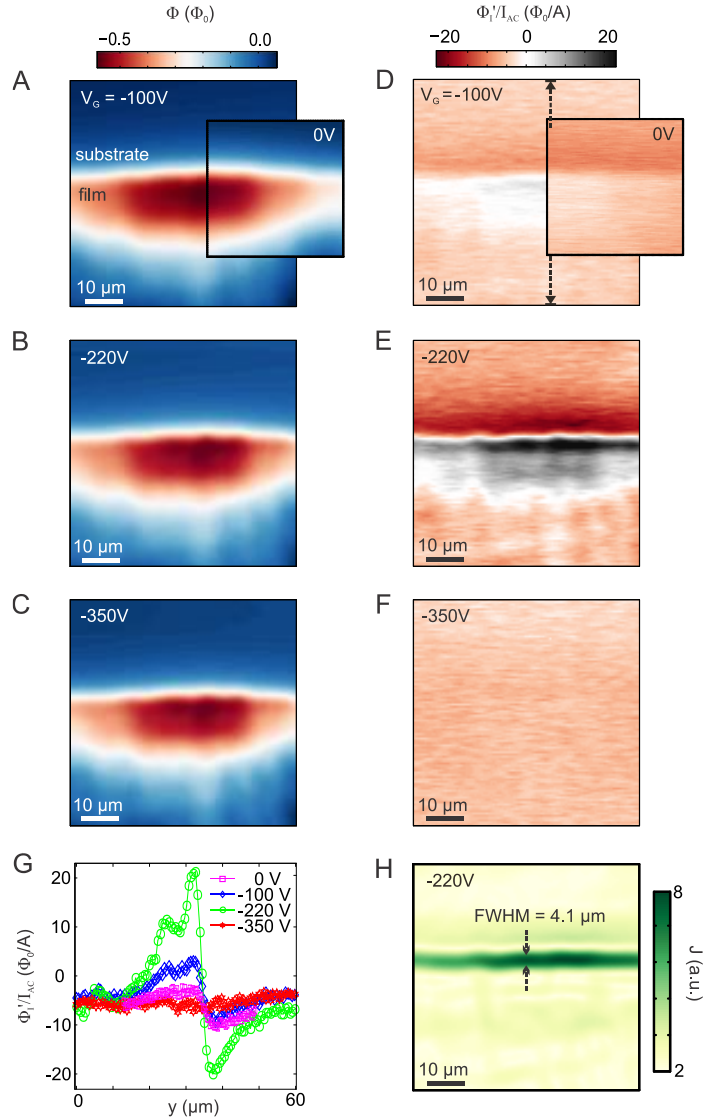

**Fig. 2 Edge current in a magnetized topological insulator appears by tuning the back gate.** (**A** to **C**) Magnetometry images along the sample edge after cooling in a uniform 30 G out-of-plane magnetic field at various backgate voltages $V_G$. (**D** to **F**) are the corresponding current magnetometry ($\Phi_I'/I_{AC}$) images. The 0-V images (**A** and **D** insets) were taken in a slightly shifted area. (**G**) Cross sections of the current magnetometry images (**D, E** and **F**) along the direction and position shown along the arrows in (**D**). (**H**) Current density for the current magnetometry image in (**E**) extracted by fast Fourier transform (see text). The arrows indicate the FWHM (full-width at half maximum) width of the edge current, which is likely resolution-limited.

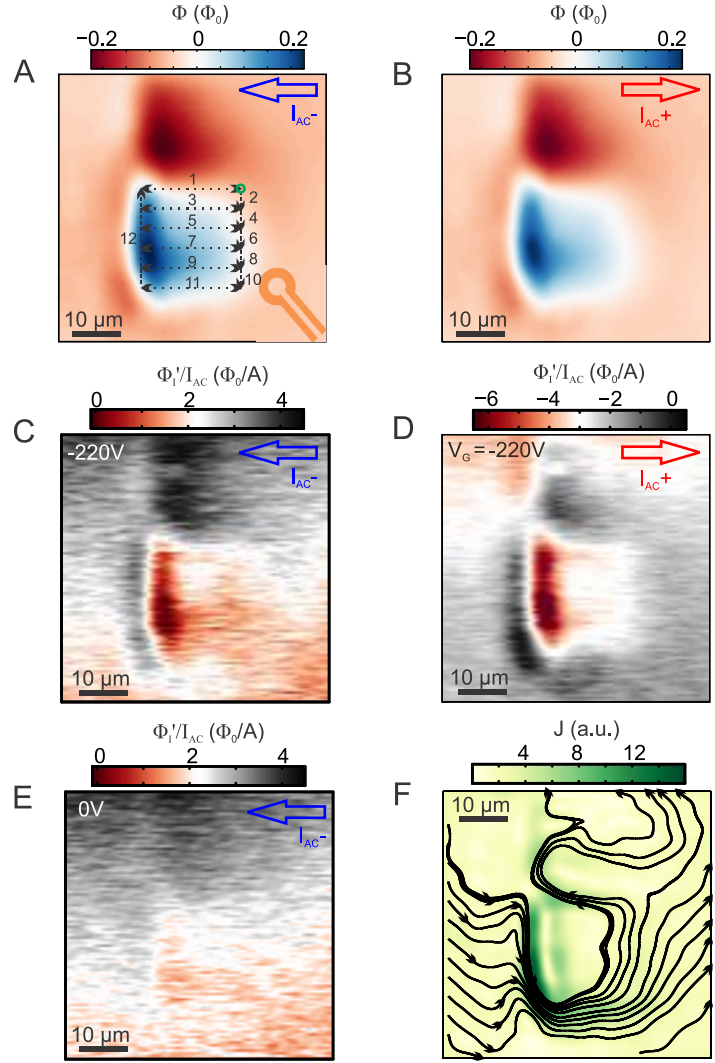

**Fig. 3 Chiral edge current emerges along the magnetic domain boundary induced by the field coil of a SQUID.** (**A**) Magnetic structure induced by applying 30-mA current to the field coil while scanning the SQUID over a 30 by 30 μm square at 12 K in the sequence indicated by the numbered dashed lines. The negative area at the top was scanned 30 μm higher with -30 mA field coil current. The field coil (orange) is sketched to scale. (**B**) The magnetometry image when the sample current bias $I_{AC}$ is reversed from (**A**) to be left-to-right. (**C** and **D**) Corresponding current magnetometry images of the magnetometry images in (A) and (B) at $V_G = -220$ V. (**E**) is the current magnetometry image with right-to-left $I_{AC}$ at $V_G = 0$ V. (**F**) Current density and current streamlines (black) extracted from the current magnetometry image in (**D**). For details about writing the magnetic structure and applying the current bias, see (31).

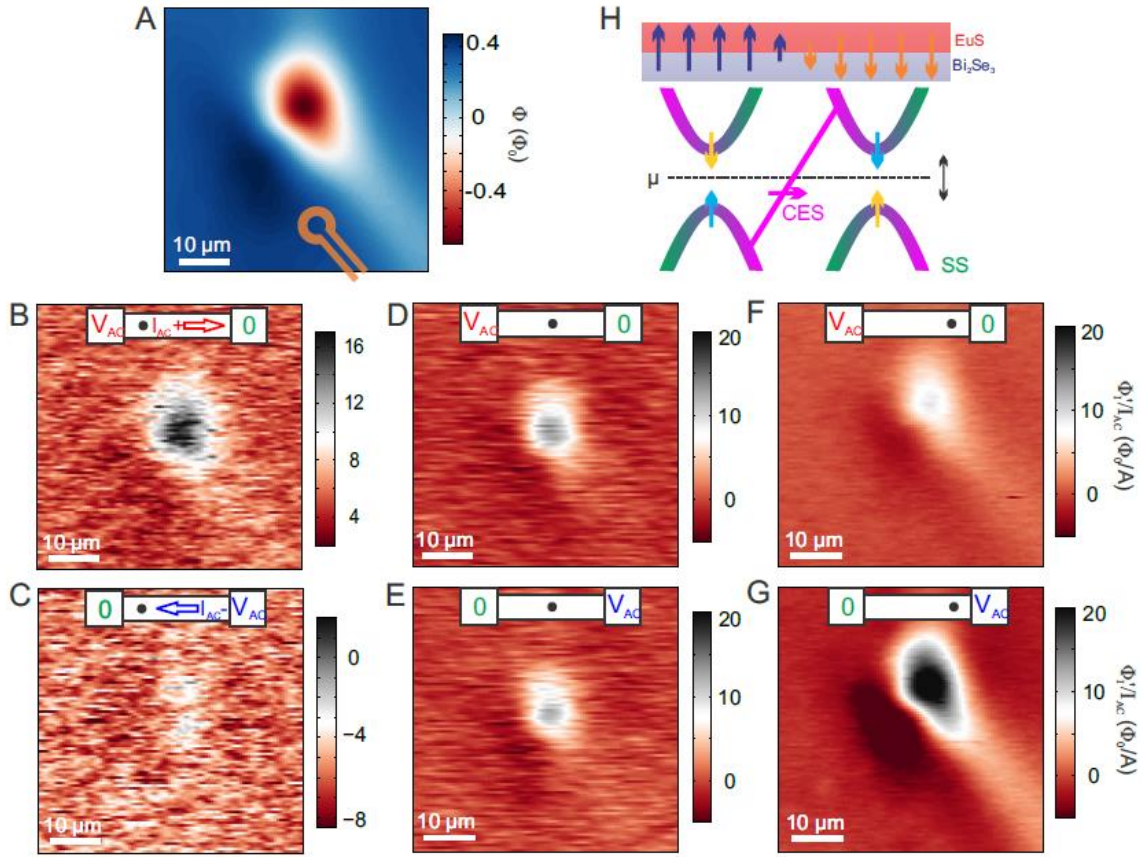

**Fig. 4 Chiral current intensity depends on electrochemical potential.** (**A**) Magnetic structure induced by the field coil when applying 30 mA of current and cooled from 15 K. The field coil (orange) is sketched to scale. (**B** to **G**) Current magnetometry images when the left side of the Hall bar is attached to a voltage lead and the right side is grounded (**B**, **D**, and **F**) versus the opposite lead configuration (**C**, **E,** and **G**). (**B**) and (**C**), (**D**) and (**E**), and (**F**) and (**G**) are from the magnetic structures on the left, middle, and right of the Hall bar, respectively (as indicated on the sketch at the top of each panel). When the magnetic structure is closer to $V_{AC}$ than to the ground, the modulation of $\mu$ is larger, yielding the stronger chiral feature in current magnetometry. (**H**) Band structure of the surface states (SS) and the chiral edge state (CES) around a magnetic domain boundary. Arrows indicate the direction of spin polarization and the color-coding of the surface states indicates how their momentum-dependent spin texture is modified by the magnetization of the top EuS layer from in-plane (green and magenta) to out-of-plane (blue and yellow). The alternating current biased source-drain voltage $V_{AC} = V_0 \sin(\omega t)$ modulates the electrochemical potential $\mu$, which is proportional to the magnitude of the edge current (see text).
17